# Gravitomagnetic Fields in Rotating Superconductors to Solve Tate's Cooper Pair Mass Anomaly

Martin Tajmar[1], Clovis de Matos[2]

[1]*Space Propulsion, ARC Seibersdorf research, A-2444 Seibersdorf, Austria*
[2]*European Space Agency - HQ, F-75015 Paris, France*
+43-50550-3142; martin.tajmar@arcs.ac.at

**Abstract.** Superconductors have often been used to claim gravitational anomalies in the context of breakthrough propulsion. The experiments could not be reproduced by others up to now, and the theories were either shown to be wrong or are often based on difficult to prove assumptions. We will show that superconductors indeed could be used to produce non-classical gravitational fields, based on the established disagreement between theoretical prediction and measured Cooper-pair mass in Niobium. Tate et al failed to measure the Cooper-pair mass in Niobium as predicted by quantum theory. This has been discussed in the literature without any apparent solution. Based on the work from DeWitt to include gravitomagnetism in the canonical momentum of Cooper-pairs, the authors published a number of papers discussing a possibly involved gravitomagnetic field in rotating superconductors to solve Tate's measured anomaly. Although one possibility to match Tate's measurement, a number of reasons were developed by the authors over the last years to show that the gravitomagnetic field in a rotating quantum material must be different from its classical value and that Tate's result is actually the first experimental sign for it. This paper reviews the latest theoretical approaches to solve the Tate Cooper-pair anomaly based on gravitomagnetic fields in rotating superconductors.

## INTRODUCTION

Present propulsion systems allow us to send robots into space and man on the moon. In the near future, nuclear propulsion systems will enable us to send humans even to Mars and probably beyond. Our technology, however, limits all our efforts to the solar system. The impossibility of sending a crew just to our next star is simply illustrated by the fact that a 40 year round-trip using the most advanced nuclear propulsion system would require a propellant mass equal to our sun (Tajmar, 2003). Therefore, any propulsion system going beyond our solar system requires a breakthrough in propulsion and power technologies. NASA launched the breakthrough propulsion physics program (Millis, 1999) to investigate some emerging theories and claims in the literature that could lead to such a breakthrough. Often, a technology capable of modifying gravity is believed to be such a breakthrough that can enable radically new propulsion systems. Although this has shown to be probably not the case (Tajmar and Bertolami, 2005), gravity control would be indeed a very powerful tool to develop new kinds of technology with possible implications for space exploration.

How can we achieve a technology that can generate gravitational fields? It has been shown already in the 1960s by Forward and others (Forward, 1961; Li and Torr, 1991; Huei, 1983; Tajmar and de Matos, 2001) that general relativity in the weak field approximation resembles a Maxwellian structure close to electromagnetism. These so-called Einstein-Maxwell equations unveil a gravitomagnetic field $B_g$ in addition to the Newtonian gravitational field $g$:

$$div\,\vec{g} = -\frac{\rho_m}{\varepsilon_g},$$

$$div\,\vec{B}_g = 0,$$

$$rot\,\vec{g} = -\frac{\partial \vec{B}_g}{\partial t},$$

$$rot\,\vec{B}_g = -\mu_g \rho_m \vec{v} + \frac{1}{c^2}\frac{\partial \vec{g}}{\partial t}.$$

(1)

These equations show that there is a Faraday-type induction law to generate gravitational fields out of time-varying gravitomagnetic fields. This is the basic principle of gravitational technologies (Forward, 1963). However, the big question is, from where do we get large gravitomagnetic fields? To give some orders of magnitude, the gravitomagnetic field generated by the whole rotating Earth is about $10^{-14}$ rad/s$^{-2}$. Gravitational fields generated out of this gravitomagnetic field can only be detected by highly specialized satellites such as Gravity-Probe B (Buchman et al, 2000). It was soon realized, that there are no means using classical matter to generate measureable gravitational fields in the laboratory (Braginski, Caves and Thorne, 1977).

In the beginning of the 1990s, both claims (Podkletnov and Nieminen, 1992, Podkletnov, 1997) and theories (Li and Torr, 1991) emerged, suggesting that superconductors can shield gravity. How does that fit into our picture of general relativity as outlined above? Well, the basic assumptions of Li et al leading to large gravitomagnetic fields in superconductors where found not to be realistic (Kowitt, 1994, Harris, 1999) and the claims of Podkletnov could not be reproduced (Li et al, 1997, Woods, Conke, Helme and Caldwell et al, 2001, Hathaway, Cleveland and Bao, 2003, Tajmar et al, 2005) so far. Nevertheless, new claims were arising (Podkletnov et al, 2003) and other theoretical venues (Modanese, 2003) were pursued to further investigate the topic.

The purpose of this paper is to show that indeed superconductivity can open up a door to generate large gravitomagnetic fields – based on a well published experiment that failed to measure the Cooper-pair mass in Niobium as predicted by quantum theory.

## THE TATE COOPER-PAIR MASS ANOMALY IN NIOBIUM

One feature of superconductivity is that the canonical momentum integrated along a close path is quantized. If the superconductor is thicker than its penetration depth (on the order of 100 nm), then the canonical momentum integral can be even set to zero. This is simply expressed by

$$\oint \vec{p}_S \cdot d\vec{l} = \oint (m\vec{v}_S + e\vec{A}) \cdot d\vec{l} = 0,$$  (2)

where $m$ and $e$ are the mass and charge of the Cooper-pair, $A$ the magnetic vector potential, and $v_s$ the speed of the Cooper-pairs. If a superconductor that was cooled down at rest ($v_s=0$) is now set into rotation (i.e. , $v_s\neq 0$, see **Fig. 1**), then it has to build up a magnetic field to still fulfill Eq. (2). One can easily transform Eq. (2) into:

$$\vec{B} = -\frac{2m}{e}\vec{\omega}.$$  (3)

This very important relation is called the London moment. It is remarkable, because a magnetic field is generated without the permeability $\mu_0$, which appears in all classical equations involving the magnetic fields. This is important as we will see later.

By accurately measuring the magnetic field of a rotating superconductor (e.g. by using a SQUID) and the angular velocity $\omega$, one can calculate the mass of the Cooper-pair (as the charge is always two times the elementary charge). This has been done for a number of superconductors (Tajmar and de Matos, 2005 and references therein), the most important result was that the Cooper-pair mass is very close to two times the electron mass independent on the material used. Tate et al performed the most accurate experiment up to now (Tate et al, 1989, 1990) revealing that the Cooper-pair mass is $m^*/2m_e = 1.000084(21)$ actually a little bit larger than two times the electron mass. This is even more a surprise as quantum theory including relativistic corrections expect the Cooper-pair mass to be a little bit smaller than two times the electron mass $m^*/2m_e = 0.999992$. The difference between experiment and theory is more than 4 sigma! This anomaly was discussed in the literature without any apparent solution (Tajmar et al, 2005 and references therein). It is even more striking that such a mass increase is simply impossible from the thermodynamic point of view.

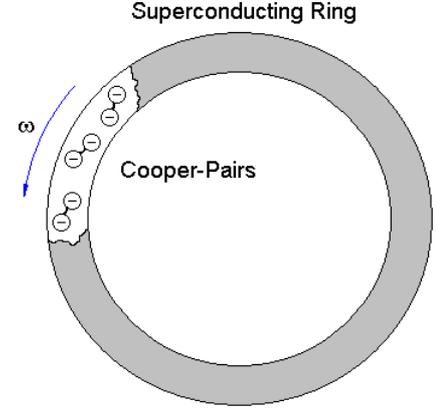

**FIGURE 1.** Rotating Superconductor.

Therefore, something must be wrong – or putting in better words – not complete. The authors recently suggested (Tajmar and de Matos, 2003, 2005) that Eq. (2) must be replaced by the full canonical momentum equation. This is not new and was first noted by DeWitt in the 1960s (DeWitt, 1966). Following out initial statements, gravitomagnetic fields are believed to be so small compared to electromagnetic ones that they can be easily neglected. But are gravitomagnetic fields inside superconductors really negligible? For non-coherent matter, the coupling between magnetic and gravitomagnetic fields can be expressed by (Tajmar and de Matos, 2001)

$$\vec{B}_g = -\frac{\mu_{0g}}{\mu_0}\frac{m}{e}\vec{B} . \tag{4}$$

Eq. (4) shows that classically, the fields are related by their mass-to-charge ratio as well as the ratio between the gravitomagnetic and magnetic permeability. As $\mu_{0g}=9.3 \times 10^{-27}$ and $\mu_0=4\pi \times 10^{-7}$, it is clear why gravitomagnetic fields can be usually neglected. However, as we have seen in Eq. (3), coherent matter produces a magnetic field without $\mu_0$, and hence, the classical coupling as written in Eq. (4) can not be applied. The full canonical momentum in Eq. (2) leads to:

$$\oint \vec{p}_S \cdot d\vec{l} = \oint \left( m\vec{v}_S + e\vec{A} + m\vec{A}_g \right) \cdot d\vec{l} = 0 , \tag{5}$$

where $A_g$ is the gravitomagnetic vector potential. Applying it to our case of a rotating superconductor, we get:

$$\vec{B} = -\frac{2m}{e}\cdot\vec{\omega} - \frac{m}{e}\cdot\vec{B}_g . \tag{6}$$

Comparing with Eq. (3), this shows that a rotating superconductor is generating a gravitomagnetic field in addition to a magnetic field. Using Tate's experimental values and the theoretical predictions (Tajmar and de Matos, 2005), we get the gravitomagnetic field that is necessary to correct Tate's result and to comply with quantum theory:

$$\vec{B}_g = 2\vec{\omega}\left(\frac{\Delta m}{m}\right) = 1.84 \times 10^{-4}\,\vec{\omega} , \tag{7}$$

where $\Delta m$ is the difference between experimental and theoretical Cooper-pair mass. If Tate's result is correct and quantum theory holds, then a rotating superconductor can indeed produce a gravitomagnetic field which is many orders of magnitude above the non-coherent matter result.

## SUPERCONDUCTIVITY AND GRAVITON MASS

We have shown so far that there is an experimental disagreement between the Cooper-pair mass and its theoretical expectation value which leaves the door open to possibly large gravitomagnetic fields. But what is the reason for such large fields in coherent-matter systems? The fact that the permeabilities are missing and that the classical coupling can not apply is a good indication but not an explanation. The answer to this question lies in the foundations of superconductivity – quantum field theory (QFT).

In modern Quantum Field Theory (QFT) superconductivity is explained in the following way: As a superconductor is passing its critical temperature, gauge symmetry is broken. This causes the photon to aquire mass via the Higgs mechanism (Ryder, 2003). The London penetration depth that we observe is then just the wavelength of the massive photon ($\lambda_L = \lambda_{Photon}$). One consequence of a massive photon is that the Maxwell equations transform into the Proca equations with two additional terms,

$$div\, \vec{E} = \frac{\rho}{\varepsilon_0} - \left(\frac{m_{Photon} c}{\hbar}\right)^2 \cdot \varphi, \qquad (8)$$

$$div\, \vec{B} = 0,$$

$$rot\, \vec{E} = -\frac{\partial \vec{B}}{\partial t},$$

$$rot\, \vec{B} = \mu_0 \rho \vec{v} + \frac{1}{c^2}\frac{\partial \vec{E}}{\partial t} - \left(\frac{m_{Photon} c}{\hbar}\right)^2 \cdot \vec{A}.$$

The authors have recently shown (De Matos and Tajmar, 2005) that by taking the rotational of the 4$^{th}$ equation and solving the differential equation, the photon mass reveals the two basic features of superconductivity: exponential shielding of electromagnetic fields (Meissner-Ochsenfeld effect) and the generation of a magnetic field by rotation (London moment with $\lambda_L = \lambda_{Photon}$):

$$B = B_0 \cdot e^{-\frac{x}{\lambda_{Photon}}} - 2\omega \frac{m}{e}\left(\frac{\lambda_{Photon}}{\lambda_L}\right)^2. \qquad (9)$$

In a typical superconductor, the photon mass is then about 1/1000 of the electron mass. If the photon is so massive, why shall the graviton in a superconductor not be massive as well? The Proca Einstein-Maxwell equations for a non-zero graviton mass lead to the following equation,

$$B_g = B_{g0} \cdot e^{-\frac{x}{\lambda_g}} - 2\omega \left(\frac{\lambda_g}{\lambda_{Lg}}\right)^2, \qquad (10)$$

where $\lambda_g$ is the graviton Compton wavelength, the second term in Eq. (10) can be interpreted as a gravitomagnetic London moment, just as we need to match Tate's Cooper-pair anomaly. The gravitomagnetic penetration depth $\lambda_{Lg}$ is defined as (de Matos and Tajmar, 2004):

$$\lambda_{Lg} = i\sqrt{\frac{1}{\mu_{0g}n_s m}} \quad , \tag{11}$$

where $n_s$ is the Cooper-pair density. Both the penetration depth as well as the graviton wavelength is a complex number, as required by the positive cosmological constant measured in our universe (Novello and Neves, 2003). This explains, if the graviton mass would be significant, why we don't observe gravitational shielding in superconductors (Li, Noever and Robertson, 1997, Tajmar and de Matos, 2005) although this was sometimes claimed (Podkletnov and Nieminen, 1992, Podekletnov, 1997, Reiss, 1999, Rounds, 1997). A complex wavelength in Eq. (10) would only lead to oscillations of the gravitomagnetic field, not to exponential shielding. Using Tate's result, we can compute the value of the graviton mass inside a Niobium superconductor as:

$$m_g = i \cdot \sqrt{\frac{\mu_{0g}n_s m^2 \hbar^2}{c^2 \Delta m}} = i \cdot 4.61 \times 10^{-55} \ kg \quad . \tag{12}$$

This is "only" 14 orders of magnitude above its accepted free-space value from the cosmological constant measurement of $i.10^{-69}$ kg (De Matos and Tajmar, 2005), but it is still a small number. In a recent assessment, Modanese (Modanese, 2003) calculated the cosmological constant inside a superconductor taking into account the contribution of the Ginzburg-Landau wave function $\psi_{GL}$ to the Lagrangian. He found that in the case of a Pb superconductor, the cosmological constant should be on the order of $10^{-39}$ m$^{-2}$. That would lead to a complex graviton mass of $i.10^{-62}$ kg, coming closer to our estimate of $i.10^{-55}$ kg in a Nb superconductor. By comparing Eqs. (7) and (10), we find that:

$$\frac{\Delta m}{m} = \left(\frac{\lambda_g}{\lambda_{Lg}}\right)^2 \quad . \tag{13}$$

Hence, the delta of mass measured by Tate is just an expression of the ratio between the graviton wavelength and its penetration depth inside a superconductor.

## CONCLUSIONS

A non-zero graviton mass in a superconductor leads to a gravitomagnetic London moment, which has the same form as the one required solving Tate's Cooper-pair mass anomaly. If Tate's measurement is correct and quantum theory holds, the Cooper-pair mass difference in her experiment is a direct sign for a graviton mass which is 14 orders of magnitude larger than its free-space value. This then explains why gravitomagnetic fields generated by rotating superconductors in fact can be much larger than the classical ones from non-coherent matter.

Although gravitomagnetic fields based on Tate's result are still small, they should be detectable within a laboratory environment. The theoretical basis outlined in this paper shall provide a basis for further experimental and theoretical research on gravitational properties of superconductors.

## ACKNOWLEDGEMENTS

The authors would like to thank Professor Jacob Biemond who first pointed their attention to J. Tate's experiments on the cooper pair mass, in private discussions.